# Towards Beyond Communication 6G Networks: Status and Challenges

Vasileios Tsekenis†, Sokratis Barmpounakis†, Panagiotis Demestichas†, Stefan Wänstedt^, Mohammad Asif Habibi*, Hans D. Schotten*, Ozgur Umut Akgul▼, Hamed Hellaoui▼, Apostolos Kousaridas▲, Milan Zivkovic‖, Panagiotis Botsinis‖, Sameh Eldessoki‖, Milan Groshev♦, Torgny Palenius§

†WINGS ICT Solutions, Greece, ^Ericsson Research, Sweden, *Rheinland-Pfälzische Technische Universität (RPTU), ▼Nokia Strategy & Technology, Finland, ▲ Nokia Strategy and Technology, Germany, ‖Apple Technology Engineering, Germany, ♦Universidad Carlos III de Madrid, §Sony Europe, Sweden

*Abstract*—Wireless communication has profoundly transformed the way we experience the world. For instance, at most events, attendees commonly utilize their smartphones to document and share their experiences. This shift in user behavior largely stems from the cellular network's capacity for communication. However, as networks become increasingly sophisticated, new opportunities arise to leverage the network for services beyond mere communication, collectively termed Beyond Communication Services (BCS). These services encompass joint communications and sensing, network as a service, and distributed computing. This paper presents examples of BCS and identifies the enablers necessary to facilitate their realization in sixth generation (6G). These enablers encompass exposing data and network capabilities, optimizing protocols and procedures for BCS, optimizing compute offloading protocols and signalling, and employing application and device-driven optimization strategies.

## I. Introductory Remarks

The evolution of wireless and mobile networks, historically focused on communications, is currently experiencing a fundamental transformation. With the advent of fifth generation (5G), the spotlight was on Internet of Things (IoT) and Edge Computing, pushing the boundaries of network capabilities beyond mere connectivity. However, these advances, while significant, still placed connectivity at the forefront [1]. As we progress towards 6G, the architecture is shifting more radically. The network is evolving beyond conventional connectivity, attempting to accommodate and support a plethora of novel services. This expansion is not just a technological advancement but a reconsideration of the network's role in innovation, and business.

Central to this evolution is the concept of BCS. The evolution of Broadband Communication Services (BCS) extends beyond traditional communication, incorporating advanced services like sensing, localization, and 'as-a-Service' offerings including computation and AI [2]. These services are transforming industries by integrating sensors, data analytics, and computation, leading to improved efficiency and productivity. The next generation of networks requires architectural adjustments to meet the varied demands of BCS applications in sectors such as logistics, manufacturing, agriculture, and healthcare. These developments involve efficient, secure, and reliable network protocols and architectures, considering data privacy, integrity, and trustworthiness. Key Performance Indicators (KPIs) and Key Value Indicators (KVIs) focus on energy efficiency, computational reliability, and effective data volume management. The enablers of Broadband Communication Services (BCS) are diverse and essential for efficient and reliable services, integrating seamlessly into existing networks. These enablers are grouped into four categories: exposure of data and network capabilities; optimization of protocols and procedures; compute offloading protocols and signaling optimization; and application and device-driven optimization [2]. As IoT and connected devices proliferate, managing large data streams becomes crucial, especially with new functionalities like advanced sensing and localization. This requires handling diverse data without compromising control plane integrity, posing challenges in network utilization, privacy, and scalability. Additionally, integrating BCS with traditional communication services demands an architecture that synergistically combines communication, computation, and sensing.

Nonetheless, addressing the diverse needs of BCS applications requires flexible and dynamic frameworks, particularly crucial for efficiently managing the wide range of devices and applications. This includes adapting registration and mobility management to ensure seamless network functionality. Furthermore, optimizing resources for various applications, especially in Industry 4.0 scenarios, presents a significant challenge. This necessitates intelligent placement of application components across the compute continuum to meet the aforementioned diverse quality of service requirements. Moreover, the challenges in data management also include the exponential increase in the data volumes. As data volumes increase with the proliferation of IoT devices and other data sources, networks must evolve to manage this data efficiently. This involves not only processing large volumes of data but also ensuring data integrity and security. The complexity of data exposure and processing in 6G networks requires robust architectural solutions that emphasize security and trust, ensuring that data protection is integral throughout its lifecycle.

To deploy BCS in 6G and beyond networks, the following critical KPIs and KVIs are needed to be considered:

**Energy efficiency/sustainability:** The energy consumption of E2E services requires optimization to minimize the peak energy consumption. These optimizations cover a multitude of aspects, including but not limited to the temporary disabling sensing transmitters and recovers that are not involved in sensing and communication operations, adjusting the sensing operation parameters, selecting a subset of sensing nodes that can provide the requested service QoS with minimum impact, or placing or relocating the applications to process the sensing data considering the E2E implications on QoS and energy consumption.

**Real time aspects:** depending on the data producers and collectors' consent, the sensing services and the data should be available to new data consumers in real time based on specific parameters. These parameters can include refresh rate, period of time, sensing KPIs and geographical location.

**Privacy:** Methods for processing, such as data cleaning or aggregation should be in the place to disguise the information whenever necessary. Data distribution and storage requirements should be agreed upon per entity and/or per information. Techniques such as data masking or encryption, or hashing can be used to secure the data while it is being stored or transmitted. To ensure that data and model privacy is protected, it is also essential to have clear policies and procedures in place that define who has access to the data, how it is collected, and how it is stored and shared.

**Bandwidth/capacity:** Bandwidth should not hinder the flow of data and model across the different network entities involved in JCAS. The bandwidth/capacity limitations should be mitigated with proper processing and proactive management.

**Resource flexibility and optimization:** providing sensing service to detect, identify and/or track one or more objects requires flexibility at the sensing and computing resources side. For example, the computational overhead that may rise from processing, fusion, or storage of data needs to be avoided.

The rest of the paper is structured as follows. In section II, we provide background on key concepts related to network beyond communication functions and data exposure and management. In Section III, we introduces the novel architectural components, protocols, signaling, and connection procedures for compute offloading. In Section IV, we discuss application- and device-driven consideration for BCS in 6G. Finally, we conclude the article in Section V.

## II. NETWORK BEYOND COMMUNICATIONS FUNCTION AND DATA EXPOSURE AND MANAGEMENT

Traditionally, networks have been primarily focused on enabling communication between devices. However, in recent years, there has been a growing recognition that networks can play a much more important role in supporting a wide range of applications and services. This is leading to the development of new network technologies and architectures that are designed to go beyond simply providing basic connectivity. The development of new network technologies and architectures is also creating new challenges in the area of data exposure and management. These challenges mainly include (a) interfaces that support data collection, (b) data processing, (c) data distribution & scaling of interfaces, (d) trust differentiation when exposing to 3rd party applications, (e) network overload on the exposed Application Programming Interfaces (APIs), (f) privacy risks, and (g) latency/performance challenges.

The afore-presented technical solutions target to optimise certain KPIs related energy consumption, bandwidth and capacity, resource flexibility and optimization, among many others. In previous studies, e.g. [2] we have presented ideas on how architecture needs to be evolved, i.e. new functions needed and new transport, to support JCAS when all measurement nodes involved are base stations. In this study we extend our work to also include UEs as measurement nodes. The design of a solution with UE involvement has, in addition to technical requirements on resolution, etc., an objective to meet SA1 requirements [3]:

- [PR 5.8.6-4] The 5G system shall be able to provide means to authorize and configure a UE for sensing operation (e.g., based on location, time, etc.) and for establishing the communication connection needed to assist the sensing service.

- [PR 5.9.6-3] The 5G system shall be able to support means to enable RAN entities and UEs to transfer sensing measurement data to sensing processing entities in the 5G system responsible for processing and aggregation of the sensing measurement data.

As depicted in the Figure 1, sensing begins when an application sends a request via the control plane (CP) to a request function, which verifies authorization before forwarding to the sensing control function. The authorization is crucial for sending back results. The sensing control sets up measurement nodes (base stations and UEs) and the processing function, also through CP. Results are transmitted to the processing function over the data plane (DP), with CP and DP separation allowing for optimal technology and capacity use. Sensing quality varies, and a UE's participation level in sensing is indicated by a quality indicator. This is necessary when the UE, connected through a serving base station, participates in sensing. Although shown separately in the figure 1, the serving base station might be co-located with the Tx or Rx base stations..

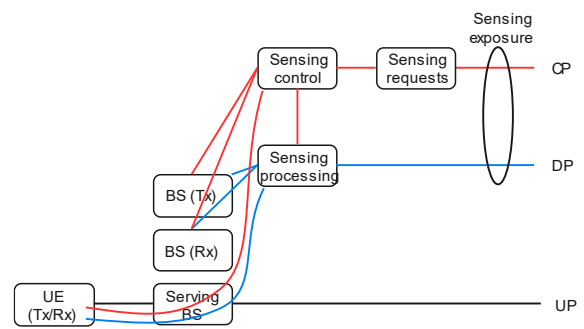

*Figure 1-: Functional architecture, network-based sensing with UE involvement for a bi-static sensing setup, ie. Tx and Rx are located in separate nodes*

The sensing request and corresponding network functions are not affected very much, or at all, by adding UE assistance. The main piece of information in a request is still the area that the application wants to investigate or sense.

The "sensing control" function (SCF) is changed a bit. The basic SCF maps translates the requested sensing area into necessary measurement nodes, i.e. determining what nodes transmit beams that cover the requested area, and provide configurations to the identified nodes to enable actual measurement. With UE assistance the SCF also finds the UEs to be involved (mapping area into relevant UEs. Note that the SCF also configures the sensing processing function (see more in the next session) with information necessary for a correct interpretation, e.g. geometry of involved nodes. The SCF may also coordinate resources between sensing and communication, e.g., sharing in time, frequency, sharing available antennas, transmission power, processing capabilities, etc., however, the final decision on scheduling is taken by the scheduler in the gNB. In a cell where resources are scarce, e.g. when the overall load is high, there will be a performance trade-off between the communication and the sensing.

As mentioned in previous deliverables [2] sensing measurements are sent to the "sensing processing" function (SPF) for processing. In most cases the SPF interprets sensing measurements applying the geometry of the involved nodes, and possible other information about the surroundings, e.g. a 3D map of buildings. Further, the SCF provides the interpretation in a format meaningful to an external receiver, i.e. answers the "question" provided in the sensing request. In case of sensor fusion information, that would be handled by the SPF. There may be use cases where raw data, measurements, reported events, etc., are forwarded to an external processing unit, however, that is not discussed any more here. This SPF also ensures that the privacy, e.g., of bystanders is protected.

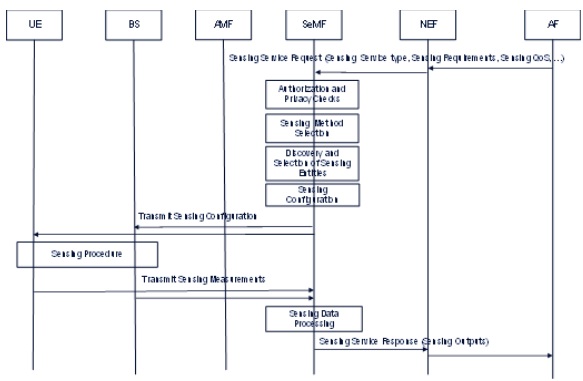

*Figure 2:- Sensing call flow where the specific sensing functionalities (e.g., SCF, SPF etc.) are placed within sensing management function (SeMF)*

To summarize, involved measurement nodes comprise sensing capable RAN entities and UEs, e.g., base stations and UEs having the capability to send and receive sensing radio signals (SeRSs) used for JCAS as well as reporting measurement results. With involved UEs comes an additional dimension as the UE may be owned by a user, why use of the UE for sensing needs to be allowed by the user. This contribution focuses on an important part in the process of involving UEs, namely, how to find an activate UEs for sensing. The sensing call flow has been fresented in Figrue 2, where the previously mentioned functionalities (e.g., SCF, SPF, etc.) are placed within a sensing management function, i.e., SeMF.

III. PROTOCOLS, SIGNALLING AND PROCEDURES

The introduction of compute offloading in the next generation of networks should be tightly integrated with the communication procedures by introducing novel architectural components for distribute computing. Moreover, to achieve an optimized Quality of Experience (QoE) for the communication and quality of computation, the utilization of communication and computation resources should be carefully designed. This imposes the following challenges in connection procedures that must be addressed:

- Discovery of the candidate compute nodes, exposure to their configuration and the definition of selection criteria
- Connection establishment between the offloading and compute nodes, their synchronization and the exchange of the computing capabilities and requirements (latency, computational load).
- Computation phase, which assumes joint compute and communication scheduling and management and transfer of the computation loads. New procedures must ensure compute service continuity and resiliency.

This section introduces the novel architectural components and connection procedures for compute offloading. It further provides the common classification of computing and communication resources and characterization of offloaded compute workloads, and relevant KPIs which enable better utilization of resources.

**A. Functional architecture for computational offloading**

Figure 3 depicts the envisioned functional architecture for computational offloading. The main introduced functional components (logical nodes) are:

- **Offloading Node**: a node that has a compute task to be offloaded.
- **Computing Node:** a node with certain processing capabilities to perform an offloaded compute task and produce compute results.
- **Compute Offload Controlling Node**: a node that collects all compute capabilities from all available Computing Nodes and makes compute offload decisions based on their current load.
- **Routing Node**: an optional node at which the compute task/compute result from Offload Node/Compute Node gets routed to one or more Computing Node(s)/Offload Node.

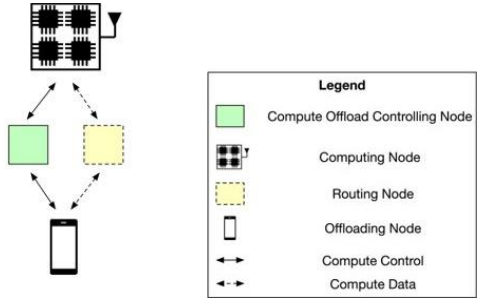

*Figure 3: Distribute compute: General functional architecture*

The physical realization of the different logical nodes in the cellular network architecture has different variants, e.g., Compute Node and Routing Node could reside in a single physical entity.

The computation offloading procedure is foreseen as a staged procedure, as illustrated in Figure 4. In the first stage (Node Discovery Phase 1), the compute node capabilities are identified. It is followed by the second stage (Node Discovery Phase 2) where the request for computation offloading is performed. Finally, the third stage (Computational Offload Procedure) comprises sending of computing tasks and receiving of compute results.

When a device, acting as an Offloading Node, decides to offload a computation, it firstly has to discover and select the candidate Compute Node(s), which would be capable of performing the requested computation while satisfying the associated KPIs. Therefore, it is important that each Compute Node estimates its execution complexity and resources (i.e., computation and storage) for given compute workload(s).

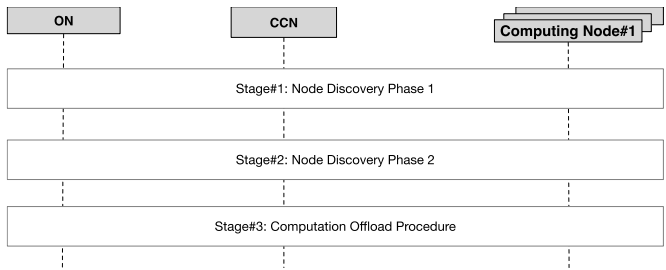

*Figure 4: Computational Offloading Procedure - High Level Flow*

### B. Common classification of compute resources, compute workloads and relevant KPIs

A common characterization of offloaded compute workloads enables better utilization of compute and communication resources. It is determined by the following requirements:

- Traffic class for specific compute task, determined by the number of nodes and the number of compute iterations between them (i.e., one-time vs multi-iteration, one-node vs multi-collaborative-nodes).
- Computation complexity (e.g., number of FLOPS and memory)
- Communication requirements (e.g., size of compute payload to be transferred, available bandwidth, supported latency)
- Precision requirement (e.g., quantization level of the compute data and operations)
- Quality of compute service classes such as latency sensitive, precision sensitive).

Furthermore, for given compute workload(s), the computing and communication resources of each Compute Node can be commonly characterized by:

- Computation capabilities (available resources)
- Memory size
- Communication capabilities/resources (available communication channels, frequency, bandwidth)

Moreover, each compute offloading/distribution operation is characterized by its cost and KPIs, such as:

- Energy consumption, both for computing operation and communication (i.e., payload transfer).
- Latency (i.e., time between the first offloaded bit until the last received result bit)
- Communication resources (i.e., required communication resources for the transfer of offloading tasks and the resulting payloads)
- Computation resources (i.e., amount of compute resources needed at the computed node)

### C. Benefits and implications of Protocols, Signalling and Procedures enabler

Through the tight integration and true convergence of communication and computing, Protocols, Signalling and Procedures enabler is likely to reduce power consumption, while keeping the device complexity low. The corresponding architectural enhancements will contribute to reducing communication and computation latency, while providing adequate data accuracy.

Moreover, compute offloading protocols, signalling and procedures enabler should bring an optimized Quality of Experience (QoE) for the communication as well as the required resiliency and quality of computation, following the "Resilence and availability" and "Support and exposure of 6G services design principles" from Hexa-X-II D2.1 deliverabe. Protocols, Signalling and Procedures enabler will have implication on RAN and core network signalling and procedures and hence new or enhanced RAN and core network procedures need to be defined. This will have an impact on related standards and regulations, primarily on RAN2 and 5G System (5GS) architecture.

Protocols, Signalling, and Procedures also face several challenges. Among them, the most critical ones are as follow:

- Discovery, control signaling protocols and procedures for computational offloading / JCAS
- true convergence of communication and computing to support resilient distributed computing.
- synchronization and coordination among offloading and computing nodes

- coordination of communication and computation resources – trade-off between comp resiliency and comm QoE
- impact of new services (sensing, localisation, etc.) on RAN interfaces and functionality

The efficient deployment of Protocols, Signalling, and Procedures are associated with the the following KPIs and KVIs: (a) Optimization of the signaling and procedures for computation, (b) Characterization of the offloading procedures, (d) Definition of generic properties of typical offloaded functions, (e) Degree of offloading, (f) Deployment (pre-deploy/adhoc), scheduling, initiator (device/network), and (g) Develop requirements and protocol scenarios and solutions that enable the introduction of Joint Communication and Sensing.

## IV. APPLICATION- / DEVICE-DRIVEN CONSIDERATIONS

The evolution of wireless and mobile networks towards 6G introduces a revolutionary change in network functionality [5], transcending the traditional focus on communication to include BCS. These services, such as sensing, enhanced localization, tracking, compute-as-a-service, and AI-as-a-Service, among others, mark a significant advancement in network capabilities. They are not just additional benefits but are integral to the network's structure, pushing the boundaries of what a network can offer to the users.

In this context, application- and device-driven optimization becomes vital. The expanding IoT environment, characterized by an exponential increase in connected devices, ranging from ambient sensors to advanced user equipment (UEs), demands a network architecture capable of managing diverse data streams. This necessitates a dual approach: on one hand, efficiently handling small data packets from power-limited sensors with their specific Quality of Service (QoS) requirements, and on the other, supporting high-volume data streams from broadband services.

The expected increase in data volume and computational load poses further challenges, particularly concerning latency, which can drastically impact performance. Efficient data processing, therefore, necessitates a reconsideration of the system's architecture, focusing on the allocation of computational tasks associated with data. This involves discovering and selecting candidate compute nodes that can perform required computations while meeting associated KPIs like latency and computational load. Therefore, a key element that emerges is the strategic placement of applications across the network. This aspect is vital in addressing the different requirements of BCS, such as latency constraints, processing needs, and data security; hence effective application placement involves the cautious positioning of applications within the network's architecture to meet their unique QoS demands. For instance, applications with high computational needs may be optimally located near powerful compute nodes, while those that need real-time processing and very low latency could be positioned closer to data generation sources. The challenge, therefore, is to integrate these placements seamlessly within the network's broader architecture, ensuring that diverse applications are supported effectively without compromising performance, security, or energy efficiency. Figure 5 provides a visual overview of this architectural complexity, portraying the need for efficient data management, security/privacy assessment, and the interplay between various domains in terms of their associated KPIs such as latency and computational load, among others.

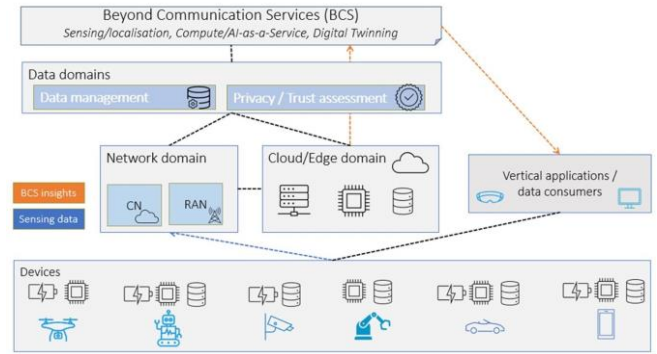

*Figure 5-: Architectural overview of Beyond Communication Services within a 6G network*

One of the critical aspects of this optimization is data management. The network must efficiently process and expose data generated by various services and applications. This includes managing the expanded data volumes, or beyond-communication data, which might require fusing at different network locations like access points for efficient processing. The design must therefore ensure scalable and coherent processing without compromising the delivery or integrity of standard control plane data. This includes developing interfaces for transferring Joint Communication and Sensing (JCAS) data to a new data plane and external entities as required.

Maintaining the integrity of the core-RAN continuum is crucial for handling data aggregation, processing, and exposure. Minimizing privacy risks and controlling data exposure becomes a significant challenge as data is cleaned and labeled across the network. Developing new architecture that prioritizes security and data protection is essential. There's a balance to be struck between data exposure and other key performance indicators. Ideally, data exposure is minimized when processing at the source, but operational needs sometimes require processing elsewhere in the network, increasing exposure risks. Balancing the benefits of exposure with privacy and security risks is necessary to optimize network performance.

Another critical aspect for power limited IoT devices. To support a good security and integrity of the data sent to and from the device, e.g. a sensor. The signalling related to the registration is today quite heavy with several messages sent fourth and back during a relatively long period between the device and the network [1], [5]. This is performed both for the initial registration but also for periodic registration and mobility related registrations. A power limited device will not be capable of doing these complex registrations but still the security and integrity of the data to and from this device is still essential. Therefore new protocols needs to be developed for this kind of devices.

One approach to enable optimal placement of BCS applications in a way to meet their QoS needs is the

Integration of Network and Compute (INC). For instance, different JCAS applications are associated with stringent requirements. On one hand, heavy computations need to be performed on the sensed data, coming from different sources, to provide with the information on their localization. On the other hand, JCAS applications could also be associated with delay-strict requirements, where the results are expected to be received in real-time (e.g., stopping a robot machine after detecting a human). In addition, although a far edge cloud is located near the end user and comes with promise of reduced communication delay, it can suffer from scarcity of compute resources which induces high processing delay. The tradeoff between network metrics and compute metrics would call for a new approach that enables the Integration of Network and Compute to perform coordinated optimization. Therefore, instead of controlling or optimizing the use of network and compute resources separately, both are considered as part of the same system and governed by common processes. This would require exposing and collecting network metrics (e.g., maximum delay and minimum throughput to an edge cloud) from network provider as well as compute metrics from cloud providers (e.g., amount of available CPUs/GPUs and memory) [5]. The availability of such metrics would allow performing an optimized decision on the placement of applications in a way to reach the requested network and compute needs.

From an architectural point of view the JCAS can be categorized in two operating modes: bistatic and monostatic. In the traditional bistatic mode, the transmitter and receiver are separated, while in the monostatic mode they are located on the same device. For example, in the traditional bistatic localization and tracking, the base station can accurately localize and track the UE. On the other hand, in the monostatic mode the UE is capable of sensing the environment and track moving objects in a so-called joint radar and communications.

A possible architectural approach in monostatic sensing is the concept of self-sensing where the mmWave radios that are integrated in the device primarily for communication purposes can be re-used to enhance the environmental sensing when the sensors experience performance degradation (i.e. in case of LiDARs with translucent material.). In the self-sensing concept a UE (i.e. robot) mounted with multiple mmWave antennas moves and scans freely the indoor environment using dedicated sensors, while the mmWave radios periodically exchange sensing information between themselves (self-sensing) by bouncing the signal in the environment, thus enabling accurate estimates of the target object/material surface.

The architectural implications of introducing the monostatic self-sensing concept and sensing data exposure are twofold. On one side, they require an architecture and protocols where the radios that perform self-sensing can exchange the sensing information with the infrastructure in a secure and efficient way, not degrading the communication performance. On the other hand, the same radios that are primarily designed and optimized for communication will need to evolve so they can support frequent sensing and addition to the communication.

To efficiently realize the abovementioned applications, the following challenges are needed to be addressed:

- Coordination of communication and computation resources – trade-off between comp resiliency and comm QoE
- BCS data consumer functions placement in terms of privacy, performance
- QoBCS-driven SLAs and dynamic network adaptability
- New architecture and protocols where the radios that perform self-sensing can exchange the sensing info with the infrasturucre in a secure and efficient way.
- New radios that can support frequent sensing.
- Exposure of application/service-driven resource control on energy limited devices
- Energy neutral devices – ambient IoT aspects (SON)

## V. CONCLUDING REMARKS

BCS represents a significant paradigm shift in the telecommunications landscape, transcending traditional communication boundaries to encompass a broader spectrum of services and applications. By integrating diverse functionalities, BCS paves the way for a seamless and unified user experience. In this paper, we explored the broader concept of BCS, defining its key characteristics and its potential impact on the telecommunications industry. We also delve into the network beyond communication function, discussing the evolution of networks beyond their traditional communication role and the emergence of new network architectures designed to support a wider range of applications and services. Then we focus on data exposure and management, examining the challenges and opportunities associated with the increasing volume and complexity of data generated and transmitted over BCS networks. Finally, we delve into the application-driven and device-driven considerations in BCS, addressing the unique performance requirements of various applications.

## VI. ACKNOWLEDGEMENT

This work has been funded by the European Union's Horizon Europe research and innovation programme under Hexa-X-II project (grant agreement No. 101095759).